\documentclass[12pt, a4paper]{article}
\usepackage{cite}
\usepackage{amsmath,amssymb}
\input{colordvi.tex}
\usepackage[dvipdfmx]{color}
\usepackage[dvipdfmx]{graphicx,hyperref}
\usepackage{comment}
\usepackage{multirow}

\hypersetup{
setpagesize=false,
 bookmarksnumbered=true,%
 bookmarksopen=true,%
 colorlinks=true,%
 linkcolor=red,
 urlcolor=blue,
 citecolor=blue,
}

\begin{document}

\begin{titlepage}

\begin{center}

\hfill UT-18-18

\vskip .75in

{\Large \bf 
Anomalous Discrete Flavor Symmetry and \\ \vspace{2mm} Domain Wall Problem
}

\vskip .75in

{\large
So Chigusa$^{(a)}$ and Kazunori Nakayama$^{(a,b)}$
}

\vskip 0.25in

$^{(a)}${\em Department of Physics, Faculty of Science,\\
The University of Tokyo,  Bunkyo-ku, Tokyo 113-0033, Japan}\\[.3em]
$^{(b)}${\em Kavli IPMU (WPI), UTIAS,\\
The University of Tokyo,  Kashiwa, Chiba 277-8583, Japan}

\end{center}
\vskip .5in

\begin{abstract}

Discrete flavor symmetry is often introduced for explaining quark/lepton
masses and mixings.  However, its spontaneous breaking leads to the
appearance of domain walls, which is problematic for cosmology.  We
consider a possibility that the discrete flavor symmetry is anomalous
under the color SU(3) so that it splits the energy levels of degenerate
discrete vacua as a solution to the domain wall problem.  We find that
in most known models of flavor symmetry, the QCD anomaly effect can only
partially remove the degeneracy and there still remain degenerate vacua.

\end{abstract}

\end{titlepage}


\renewcommand{\thepage}{\arabic{page}}
\setcounter{page}{1}
\renewcommand{\thefootnote}{\#\arabic{footnote}}
\setcounter{footnote}{0}

\section{Introduction}

Discrete flavor symmetry is often introduced in order to explain the observed patterns of neutrino masses and mixings.
Depending on the choice of flavor symmetry and its way of spontaneous breaking, various predictions have been obtained (for reviews, see Refs.~\cite{Altarelli:2010gt,Ishimori:2010au,King:2014nza,King:2017guk}).
However, spontaneous breaking of a discrete symmetry leads to the formation of domain walls~\cite{Vilenkin:2000jqa}.
Once formed, domain walls are topologically stable and come to dominate the universe, which is a disaster for cosmology.
There are several solutions.
First, if inflation happens after the discrete symmetry breaking, domain walls are inflated away and no domain walls are left in our observable universe
and hence there is no problem.
Second, if the flavor symmetry is not exact but only an approximate one, domain walls are unstable and collapse at some instance~\cite{Vilenkin:1981zs}.

Whether the first option works or not depends on the energy scale of
inflation, that of flavor symmetry breaking, and the reheating
temperature.  If the flavor symmetry breaking scale is sufficiently
larger than the inflation scale and maximum temperature of the universe,
the flavor symmetry may not be restored after inflation.  However, most
concrete models introduce supersymmetry (SUSY)~\cite{Martin:1997ns} in
order to simplify the interaction among various flavon fields and there
often appear flat directions in the scalar potential in
SUSY~\cite{Altarelli:2005yx}.  These flat directions obtain masses of
soft SUSY breaking, which is usually much lower than the flavor symmetry
breaking scale.  In this case, we need an inflation scale lower than the
SUSY breaking scale (e.g. $\mathcal O({\rm TeV})$ for solving the
hierarchy problem) to solve the domain wall problem.  Such a low-scale
inflation is not excluded, but rather unlikely.  Thus we do not pursue
this option in this paper.

As for the second option, it is always possible to introduce small
explicit symmetry breaking terms by hand unless the discrete symmetry is
a remnant of some gauge symmetry.  But there are so many ways to
introduce such breaking terms, which are not under control and may
(partly) destroy original motivation to consider flavor symmetry to
explain observed data.  Thus we restrict ourselves to a special case:
the discrete flavor symmetry is exact at the classical level and only
broken by the quantum anomaly.  The concept is the same as the
Peccei-Quinn (PQ) symmetry for solving the strong CP
problem~\cite{Peccei:1977hh,Kim:1986ax}.

The role of anomaly (in particular, anomaly under the color SU(3)) for
Abelian discrete symmetry to solve the domain wall problem was explored
in Ref.~\cite{Preskill:1991kd} and further applied to some concrete
models in Refs.~\cite{Dine:1993yw, Ibe:2004gh, Riva:2010jm,
Moroi:2011be, Hamaguchi:2011nm, Hamaguchi:2011kt}.  On the other hand,
the anomaly of general (non-)Abelian discrete groups was extensively
studied in Refs.~\cite{Ibanez:1991pr, Ibanez:1991hv, Banks:1991xj,
Araki:2007zza, Araki:2008ek, Luhn:2008sa, Chen:2013dpa, Chen:2015aba}.
In the context of model-building, non-anomalous discrete symmetry is
often used in order to embed the discrete symmetry into some continuous
gauge group.  Our attitude is completely different: we want to make use
of anomalous discrete flavor symmetry to find models without domain wall
problem.

In Sec.~\ref{sec:gen} we summarize the structure of the discrete anomaly
in general and examine all the non-Abelian discrete groups listed in
Ref.~\cite{Ishimori:2010au} to see whether the anomalous breaking of the
symmetry is allowed or not. We will find that none of them can be
completely anomalous and hence the domain wall problem is not solved
solely by the anomaly effect.  In Sec.~\ref{sec:ex} we will check this
statement with some explicit examples.  We conclude in
Sec.~\ref{sec:conc}.

\section{Discrete anomaly and degenerate vacua}  \label{sec:gen}

\subsection{Discrete anomaly} 

The anomaly of general non-Abelian discrete groups has been analyzed in
Refs.~\cite{Araki:2007zza, Araki:2008ek, Luhn:2008sa, Chen:2013dpa,
Chen:2015aba} by using the Fujikawa
method~\cite{Fujikawa:1979ay,Fujikawa:1980eg}. It was pointed out that
the measure of the path integral necessarily transforms as a
one-dimensional representation of the discrete group and hence perfect
groups, which do not have non-trivial one-dimensional representations,
are anomaly-free~\cite{Chen:2013dpa, Chen:2015aba}.  It means that we
need to only care about the Abelian subgroups, $Z_n$, of the non-Abelian
discrete groups.

Let us consider a non-Abelian discrete group $D$. What we seek is an
anomaly of the $D$-SU(3)-SU(3) type, with SU(3) being the standard model
QCD gauge group, since it is the only one that can significantly change
the potential to solve the domain wall problem through the QCD instanton
effect. If there is another strong hidden gauge interaction and a
fermion charged under both this strong gauge group and the discrete
group, it can also be important. Taking this case into account, we
consider an anomaly of the $D$-SU($N$)-SU($N$) type.

Let us suppose that there is a set of left-handed chiral fermions $\psi$
that transforms as a representation ${\bf R}$ of the gauge group SU($N$)
and also transforms as
\begin{align}
	\psi \to U_\psi(u) \psi,
\end{align}
under an element $u$ of the discrete group $D$.  Here, $U_\psi(u)$ is a
matrix representation of $u$. Then we will have a following Lagrangian
term
\begin{align}
	\delta \mathcal L = A \frac{2\pi}{n} \frac{1}{32\pi^2} F_{\mu\nu}^a \widetilde F^{\mu\nu a},   \label{L_anom}
\end{align}
where $F_{\mu\nu}^a$ and $\widetilde F^{\mu\nu a}$ denote the field
strength of the SU($N$) gauge and its dual respectively, $n$ is a
minimum integer such that $u^n={\bf 1}$ (hence $u$ generates a $Z_n$
subgroup of $D$).  The coefficient $A$ is given by
\begin{align}
	A = \sum_\psi q_\psi\, 2\ell_\psi\left({\bf R}\right),
\end{align}
where $\ell_\psi\left({\bf R}\right)$ denotes the Dynkin index of the
representation ${\bf R}$, which is $1/2$ for the fundamental
representation of SU($N$). Here $q_\psi$ is a ``charge'' of $\psi$ under
$Z_n$ rotation, defined by
\begin{align}
	q_\psi = \frac{n}{2\pi i}\ln\det U_\psi(u).  \label{charge}
\end{align}
If $\psi$ transforms as a one dimensional representation of $D$, we have
$U_\psi = e^{2\pi i q_\psi/ n}$. The anomaly-free condition is given by
\begin{align}
	A \equiv 0 ~~~{\rm mod}~~~n.  \label{A=0}
\end{align}
If this condition is violated, the discrete group $D$ is anomalous under
the SU($N$). Then the anomaly (\ref{L_anom}) has a physical meaning that
several vacua connected by the discrete transformation have different
energy due to the instanton effect. Thus it can serve as a bias that
renders domain walls unstable. It is completely analogous to the PQ
solution to the strong CP problem: if we would consider global U(1)
instead of discrete group $D$ and if it is anomalous under the QCD, the
flat potential along the U(1) direction (axion) is lifted up by the
instanton effect.

If there are several discrete groups $D$ under which the Lagrangian is
classically invariant, as is often the case for models with flavor
symmetry, we have
\begin{align}
	\delta \mathcal L = \sum_D\left({A_D \frac{2\pi}{n_D}}\right) \frac{1}{32\pi^2} F_{\mu\nu}^a \widetilde F^{\mu\nu a}.  \label{L_anomaly}
\end{align}
If the condition (\ref{A=0}) is satisfied for each discrete group, the
theory is obviously anomaly-free.  If this condition is violated, the
discrete symmetry is anomalous. Still, however, it is possible that
there remain some subgroups of the original discrete symmetry free from
anomalies. In order to ensure the non-existence of domain walls, there
must be no remnant discrete symmetry.\footnote
{
The underlying assumption here is that vacuum expectation values (VEVs)
of the flavon fields spontaneously break the discrete symmetry $D$
completely. If some subgroup is not spontaneously broken by VEVs of the
flavons, what we need for the non-existence of domain walls is to
explicitly break only the spontaneously broken part of the discrete
symmetry.
}
For this purpose, we consider all the elements of $Z_n$ rotations:
\begin{align}
	\psi \to \left( U_\psi(u) \right)^k \psi,
\end{align}
with $k=0,\dots, n-1$. Then the anomaly coefficient becomes
\begin{align}
	\mathcal A = \sum_D k_D\frac{A_D}{n_D},   \label{anom_coef}
\end{align}
with $k_D = 0,\dots, n_D-1$. In order to remove all the degeneracy of
the vacua, $\mathcal A$ should not be an integer for any choice of
$k_D$, except for the trivial one (i.e., all $k_D$ are equal to
zero). Otherwise, there remain discrete degenerate vacua.

\begin{table}[t]
 \begin{center}
  \def\arraystretch{1.2}
  \begin{tabular}{|c|c|}
  \hline
  Discrete group & $\det U_\psi$ of representations \\ \hline
  \multirow{2}{*}{$S_3 \cong Z_3 \rtimes Z_2$} & $\textbf{1}' \to (1,
       -1)$ \\
   & $\textbf{2} \to (1, -1)$ \\ \hline
  \multirow{3}{*}{$S_4 \supset Z_3, Z_4$} & $\textbf{1}' \to (1, -1)$ \\
   & $\textbf{2} \to (1, -1)$ \\
   & $\textbf{3} \to (1, -1)$ \\ \hline
  \multirow{2}{*}{$A_4 \cong (Z_2 \times Z_2) \rtimes Z_3$} &
       $\textbf{1}' \to (1, 1, \omega)$ \\
  & $\textbf{1}'' \to (1, 1, \omega^2)$ \\ \hline
  $A_5 \supset Z_2,Z_3$ & None \\ \hline
  \multirow{4}{*}{$T' \supset Z_3, Z_4$} & $\textbf{1}' \to (\omega, 1)$
       \\
   & $\textbf{1}'' \to (\omega^2, 1)$ \\
   & $\textbf{2}' \to (\omega^2, 1)$ \\
   & $\textbf{2}'' \to (\omega, 1)$ \\ \hline
  \end{tabular}
 \end{center}
 \caption{List of frequently used discrete groups and their
 representations which can contribute to the anomaly.  In the left
 column, discrete groups and their Abelian subgroups are shown.  In the
 right column, $\det U_\psi$ (see Eq.~(\ref{charge})) of relevant
 representations under the Abelian subgroups are shown.  Here we defined
 $\omega = e^{2\pi i /3}$ and $\rho = e^{2\pi i /N}$.}
 \label{130301_23Aug18}
\end{table}

\begin{table}[t]
 \begin{center}
  \def\arraystretch{1.2}
  \begin{tabular}{|c|c|}
   \hline
   Discrete group & $\det U_\psi$ of representations \\ \hline
   \multirow{4}{*}{$D_{2N} \cong Z_{2N} \rtimes Z_2$} & $\textbf{1}_{+-}
       \to (-1, 1)$ \\
   & $\textbf{1}_{-+} \to (-1, -1)$ \\
   & $\textbf{1}_{--} \to (1, -1)$ \\
   & $\textbf{2}_k \to (1, -1)~~(1 \leq k < N)$ \\ \hline
  \multirow{2}{*}{$D_{2N-1} \cong Z_{2N-1} \rtimes Z_2$} &
       $\textbf{1}_{-} \to (1, -1)$ \\
   & $\textbf{2}_k \to (1, -1)~~(1 \leq k < N)$ \\ \hline
  \multirow{4}{*}{$Q_{4N} \supset Z_{4N}, Z_4$} & $\textbf{1}_{+-} \to
       (-1, 1)$ \\
   & $\textbf{1}_{-+} \to (-1, -1)$ \\
   & $\textbf{1}_{--} \to (1, -1)$ \\
   & $\textbf{2}_k \to (1, -1)~~(k={\rm even})$\\ \hline
   \multirow{4}{*}{$Q_{4N+2} \supset Z_{4N+2}, Z_4$} &
       $\textbf{1}_{+-} \to (-1, i)$ \\
   & $\textbf{1}_{-+} \to (-1, -i)$ \\
   & $\textbf{1}_{--} \to (1, -1)$ \\
   & $\textbf{2}_k \to (1, -1)~~(k={\rm even})$\\ \hline
  \end{tabular}
 \end{center}
 \caption{Continuation from Table~\ref{130301_23Aug18}.}
 \label{140128_24Aug18}
\end{table}

\begin{table}[t]
 \begin{center}
  \def\arraystretch{1.2}
  \begin{tabular}{|c|c|}
   \hline
   Discrete group & $\det U_\psi$ of representations \\ \hline
   \multirow{3}{*}{$\Sigma(2N^2) \cong (Z_N \times Z_N) \rtimes Z_2$} &
       $\textbf{1}_{+n} \to (\rho^n, \rho^n, 1)~~(1\leq n < N)$ \\
   & $\textbf{1}_{-n} \to (\rho^n, \rho^n, -1)~~(0\leq n < N)$ \\
   & $\textbf{2}_{p,q} \to (\rho^{p+q}, \rho^{p+q}, -1)~~(0 \leq p,q <
       N)$ \\ \hline
   $\Delta (3N^2) \cong (Z_N \times Z_N) \rtimes Z_3$ & $\textbf{1}_1 \to
       (1, 1, \omega)$ \\
   ($N/3 \neq$ integer) & $\textbf{1}_2 \to (1, 1, \omega^2)$ \\ \hline
   $\Delta(3N^2) \cong (Z_N \times Z_N) \rtimes Z_3$ &
       \multirow{2}{*}{$\textbf{1}_{k,l} \to (\omega^l, \omega^l,
       \omega^k)~~(0\leq k,l < N)$} \\
   ($N/3 = $ integer) & \\ \hline
   \multirow{2}{*}{$T_7 \cong Z_7 \rtimes Z_3$} & $\textbf{1}_1 \to (1,
       \omega)$ \\
   & $\textbf{1}_2 \to (1, \omega^2)$ \\ \hline
   \multirow{3}{*}{$\Sigma(81) \cong (Z_3 \times Z_3 \times Z_3) \rtimes
   Z_3$} & $\textbf{1}_l^k \to (\omega^k, \omega^k, \omega^k,
       \omega^l)~~(0\leq k,l < 3)$ \\
   & $\textbf{3}_a \to (\omega, \omega, \omega, 1)~~(a=A,B,C)$ \\
   & $\textbf{3}_{\bar{a}} \to (\omega^2, \omega^2, \omega^2,
       1)~~(\bar{a}=\bar{A},\bar{B},\bar{C})$ \\ \hline
   \multirow{3}{*}{$\Delta (54) \cong (Z_3 \times Z_3) \rtimes (Z_3
   \rtimes Z_2)$} & $\textbf{1}_{-} \to (1, 1, 1, -1)$ \\
   & $\textbf{2}_k \to (1, 1, 1, -1)~~(1 \leq k \leq 4)$ \\
   & $\textbf{3}_{1(k)} \to (1, 1, 1, -1)~~(k=1,2)$ \\ \hline
  \end{tabular}
 \end{center}
 \caption{Continuation from Table~\ref{140128_24Aug18}.}
 \label{131455_23Aug18}
\end{table}

From the above discussion, it is obvious that we need at least one
representation for each $Z_n$ subgroup of $D$ which transforms
non-trivially under $Z_n$.  In Table~\ref{130301_23Aug18},
Table~\ref{140128_24Aug18} and Table~\ref{131455_23Aug18}, we show a
list of discrete groups frequently used in flavor models and their
representations that take some non-trivial values of $\det U_\psi$ (see
Eq.~(\ref{charge}))~\cite{Ishimori:2010au}.  Representations with $\det
U_\psi=1$ or $q_\psi=0$ for all $Z_n$ subgroups are omitted since they
do not contribute to the anomaly.  For example, the discrete group
$S_3$, which is isomorphic to $Z_3 \rtimes Z_2$, possesses a non-trivial
singlet expression $\textbf{1}'$, which has a charge $q_\psi = 1$ under
$Z_2$, but is not charged ($q_\psi = 0$) under $Z_3$.  From the tables,
we can see that some of the groups (such as $A_4$) have an Abelian
subgroup ($Z_2 \times Z_2$ in the case of $A_4$), which do not transform
any fermion non-trivially.  Then we can see that such subgroups cannot
be anomalous.  On the other hand, there are also examples such as
$D_{2N}$, which do not have a $Z_n$ subgroup under which all the
fermions are trivial. Note also that $A_5$, which is perfect, do not
have non-trivial representations as we mentioned above.

Possible candidates of the groups that may be completely anomalous are
$D_{2N}$, $Q_{4N}$, $Q_{4N+2}$, $\Sigma(2N^2)$, $\Delta(3 N^2)$ ($N/3=$
integer), and $\Sigma(81)$.  For $D_{2N}$, we see that all non-trivial
representations have charge $q_\psi=N$ under $Z_{2N}$ and hence there
must remain unbroken $Z_N$ symmetry.  Similarly, for $Q_{4N}$ and
$Q_{4N+2}$, $Z_{2N}$ and $Z_{2N+1}$ subgroup cannot be anomalous,
respectively.  For $\Sigma(2N^2)$, all relevant representations have the
same charge under two $Z_N$ subgroups. Thus one linear combination of
each $Z_N$ remains unbroken. Similar argument also applies to $\Delta(3
N^2)$ ($N/3=$ integer) and $\Sigma(81)$.  After all, all the groups
listed here cannot be completely broken solely by the anomaly effect,
and hence the domain wall problem is not solved.

In Sec.~\ref{sec:ex}, we will explicitly show that vacuum degeneracy is
partially removed by the QCD anomaly effect, but it is not enough to
completely solve the domain wall problem. In particular, we focus on the
examples with $A_4$ and $D_4$.

\subsection{The case with global U(1)} 

Discrete flavor symmetry may naturally explain observed neutrino masses
and mixings, but often it can not explain the hierarchical mass
structure of the quarks and charged leptons.  In many concrete models of
flavor symmetry, a global U(1) symmetry is introduced in order to
explain quark/lepton masses~\cite{Altarelli:2010gt}.  Hierarchical
Yukawa couplings may be explained through Froggatt-Nielsen (FN)
mechanism~\cite{Froggatt:1978nt}.  In the FN mechanism, quarks and/or
leptons are assumed to have charges under global U(1), called U(1)$_{\rm
FN}$, and the VEV of a flavon field with U(1)$_{\rm FN}$ charge
naturally explains the hierarchical mass structure.  Even if the
U(1)$_{\rm FN}$ is exact at the classical level, it may have a QCD
anomaly. In such a case, we can identify the U(1)$_{\rm FN}$ as the PQ
symmetry to solve the strong CP problem~\cite{Wilczek:1982rv,
Albrecht:2010xh, Ahn:2014gva, Nomura:2016nfi, Ema:2016ops,
Calibbi:2016hwq}.

Let us consider the case where a global U(1) is introduced in addition
to the discrete flavor group $D$, and suppose that both the U(1) and $D$
is anomalous under the QCD.  Then the anomaly coefficient
(\ref{anom_coef}) becomes
\begin{align}
	\mathcal A = \frac{\theta}{2\pi n_{U(1)}} + \sum_D k_D\frac{A_D}{n_D},
\end{align}
where $\theta = 0\sim 2\pi$ is the U(1) rotation angle, corresponding to
the Nambu-Goldstone (NG) boson, and $n_{U(1)}$ is a non-zero integer
determined by the U(1) charge of quarks.  This $\theta$ is a dynamical
field and it is dynamically relaxed to the potential minimum so that
$\mathcal A = 0$ due to the QCD instanton effect and solves the strong
CP problem.  In this case, the QCD anomaly effect works as a solution to
the strong CP problem, but it has nothing to do with the domain wall
problem associated with the spontaneous breaking of the discrete group
$D$~\cite{Preskill:1991kd}.  Even if it is possible to assign the quark
charges so that the discrete group $D$ is completely anomalous under the
QCD, such an anomaly is canceled by the shift of the NG boson and the
discrete minima reappear, independently of the value of $n_{U(1)}$.
Thus we do not consider a global U(1) hereafter.

\section{Examples} \label{sec:ex}

\subsection{$A_4$ model}

The so-called tri-bimaximal neutrino mixing
pattern~\cite{Harrison:2002er} can be achieved by models with $A_4$
flavor symmetry~\cite{Ma:2001dn,Altarelli:2005yx}.  However, the
discovery of $\theta_{13}$~\cite{An:2012eh} requires modification to the
tri-bimaximal mixing.  Introducing additional scalar $\xi'$ which
transforms as ${\bf 1'}$ under $A_4$ can explain small but non-zero
$\theta_{13}$~\cite{Shimizu:2011xg} (see also Ref.~\cite{Kang:2018txu}).
Table~\ref{table:A4} summarizes charge assignments of quarks, leptons,
Higgs and flavon fields under $A_4$ and additional $Z_N'$.\footnote{ In
order to distinguish the $Z_3$ subgroup of $A_4$ from the additional
discrete Abelian group, we use the notation $Z_N'$ for the latter.  } We
sometimes use the notation e.g. $d^c_i$ with $i=1,2,3$ corresponding to
$d^c,s^c,b^c$ respectively, and so on.

The Yukawa interactions responsible for lepton mass matrices are given as\footnote{
	We introduce up- and down-type Higgs doublets keeping SUSY setup in mind, although the most discussion in this paper does not require SUSY.
}
\begin{align}
	\mathcal L = &y_e \frac{e^c (L \phi_l) H_d}{M} + y_\mu \frac{\mu^c (L \phi_l)' H_d}{M} + y_\tau \frac{\tau^c (L \phi_l)''H_d}{M} \nonumber\\
	&+ y_\nu\frac{(\phi_\nu LL) H_uH_u}{M^2} + y_\xi\frac{\xi(LL) H_uH_u}{M^2} + y_{\xi'}\frac{\xi'(LL)'' H_uH_u}{M^2} + {\rm h.c.},
\end{align}
where $(\cdots)$, $(\cdots)'$ and $(\cdots)''$ denote ${\bf 1}$, ${\bf
1'}$ and ${\bf 1''}$ product of $A_4$ triplets, respectively.  After
having VEVs of
\begin{align}
	\left<\phi_l\right> = (v_l,0,0),~~~~~~\left<\phi_\nu\right> = (v_\nu,v_\nu,v_\nu),~~~~~~
	\left<\xi\right> = v_\xi,~~~~~~\left<\xi'\right>=v_{\xi'},  \label{A4_min}
\end{align}
it can reproduce the charged lepton masses, neutrino masses and mixings
as discussed in detail in Ref.~\cite{Shimizu:2011xg}, so we do not
repeat here.

As for the quark sector, we can write the Lagrangian as follows:
\begin{align}
	\mathcal L = y^{ij}_d d^c_i Q_j H_d + y^{Ij}_u  u^c_I Q_j H_u + y_u^{1j}\frac{\xi' u^c_1 Q_j H_u}{M} + {\rm h.c.},
\end{align}
where $i,j=1,2,3$ and $I=2,3$.  After having VEV of $\xi'$, this reduces
to completely general Yukawa matrices for up and down quarks, so clearly
there are enough degrees of freedom to reproduce observed quark masses
and mixings.  The reason for this particular choice of charge in the
quark sector is that $A_4$ and $Z_N'$ become anomalous under the QCD
since only one of the right-handed up-quarks is charged under these
discrete groups.  This is just one simple example and there are many
other possible choices.

\begin{table}
\begin{center}
\begin{tabular}{|c|cccc|cccc|cc|cccc|} \hline
    ~        & $Q_i$ & $u^c$ & $c^c,t^c$ & $d^c_i$ & $L_i$ & $e^c$ & $\mu^c$ & $\tau^c$ & $H_u$ & $H_d$ & $\phi_l$ & $\phi_\nu$ & $\xi$ & $\xi'$ \\ \hline
 $A_4$   & ${\bf 1}$ & ${\bf 1''}$ & ${\bf 1}$ & ${\bf 1}$ & ${\bf 3}$ & ${\bf 1}$ & ${\bf 1''}$ & ${\bf 1'}$ & ${\bf 1}$ & ${\bf 1}$ & ${\bf 3}$& ${\bf 3}$& ${\bf 1}$& ${\bf 1'}$    \\ 
 $Z_N'$  &$1$ & $\omega_N^2$ & $1$ & $1$ & $\omega_N$ & $\omega_N^{-1}$ & $\omega_N^{-1}$ & $\omega_N^{-1}$ & $1$ & $1$ & $1$& $\omega_N^{-2}$& $\omega_N^{-2}$& $\omega_N^{-2}$    \\ \hline
\end{tabular}
\caption{Charge assignments under $A_4$ and $Z_N'$ for quarks, leptons
and various Higgs and flavon fields. Here $\omega_N = e^{2\pi i /N}$ is
an element of $Z_N'$ rotation.}  \label{table:A4}
\end{center}
\end{table}

Now let us explore the vacuum structure of the model at the classical
level.  Details of the shape of the entire scalar potential are not
needed for our purpose. The only information we need is that
(\ref{A4_min}) is one of the potential minima.
Any transformation of (\ref{A4_min}) by the $A_4$ and $Z_N'$ group
element is also a vacuum.
In the following, we see that there should be $12N$ ($N=$ odd) or $6N$
($N=$ even) discrete vacua with degenerate energy (at the classical
level).  First, for fixed $Z_N'$ angle, we find 12 vacua consisting of
some combinations of the following $\left<\phi_l\right>$ and
$\left<\phi_\nu\right>$ where
\begin{align}
	\left<\phi_l\right> = v_l(1,0,0),~~~\frac{v_l}{3}(-1,2,2),~~~\frac{v_l}{3}(-1,2\omega,2\omega^2),~~~\frac{v_l}{3}(-1,2\omega^2,2\omega),
	\label{phi_l}
\end{align}
where $\omega \equiv e^{2\pi i/3}$ and
\begin{align}
	\left<\phi_\nu\right> =\pm v_\nu(1,1,1),~~~\pm v_\nu(1,\omega,\omega^2),~~~\pm v_\nu(1,\omega^2,\omega).
	\label{phi_nu}
\end{align}
For each vacuum, $\left<\phi_\nu\right>$ transformed by the $Z_N'$
symmetry is also a vacuum and hence we have $12N$ ($N=$ odd) or $6N$
($N=$ even) vacua in total.\footnote
{
Since the ``charge'' of the flavon $\phi_\nu$ under the $Z_N'$
symmetry is $-2$, there are only $N/2$ different transformations when
$N$ is even: rotations by an overall factor $\omega_N^n$ with $n=0, 2,
\dots, N-2$.  On the other hand, there are $N$ different
transformations with $n=0,1,\dots,N-1$ when $N$ is odd.
}
The existence of $12$
discrete vacua (for fixed $Z_N'$ phase) reflects the fact that all the
$A_4$ symmetry is spontaneously broken by the VEVs of flavon
fields. (If some subgroup of the $A_4$ symmetry remains unbroken, some
part of the 12 vacua may coincide with each other.)

The degeneracy of these vacua at the classical level is partly lifted by
the effect of the QCD anomaly. Since the up-quark transforms
non-trivially under $A_4$, in particular under its $Z_3$ subgroup, vacua
which are connected by the $T$ transformation will actually have a
different energy of order $\sim (m_\pi f_\pi)^2$ with $m_\pi$ and
$f_\pi$ being the pion mass and decay constant, respectively.\footnote {
We follow the notation of Ref.~\cite{Altarelli:2010gt} for the $A_4$
group element: $S^2=1$ and $T^3=1$ in the basis where $T$ is
diagonal. See App.~\ref{app:A4}.  } Similarly, the $Z_N'$ symmetry is
also anomalous and vacua connected by the $Z_N'$ transformation are also
lifted.  The anomaly coefficient (\ref{anom_coef}) is
\begin{align}
	\mathcal A = k_1\frac{2}{3} + k_2\frac{2}{N},
\end{align}
where $k_1=0,1,2$ and $k_2=0,\dots, N-1$. For $N=5$, for example,
there are no solutions of $(k_1,k_2)$ that make $\mathcal A$ integer
except for the trivial one $k_1=k_2=0$. Thus $Z_3\times Z_N'$ may be
completely broken by the QCD anomaly and vacuum degeneracy associated
with these groups is solved.  However, the QCD anomaly has no effects
on the $Z_2 \times Z_2$ subgroup of $A_4$, since both ${\bf 1'}$ and
${\bf 1''}$ do not transform under this subgroup. Explicitly, we can
find vacua connected by $S, TST^2, T^2ST$ transformation from one
vacuum and they have degenerate energy. Therefore there exist stable
domain walls associated with them.  This argument does not depend on
the structure of the (one of the) vacuum (\ref{A4_min}). The
discussion is completely parallel for more general VEVs of
flavons. After all, the QCD anomaly of the discrete group is not
enough to solve the domain wall problem in this class of models.\footnote{
	In a concrete setup of the flavon sector~\cite{Altarelli:2005yx}, there is a flat direction in the scalar potential whose VEV spontaneously breaks $Z_3 \subset A_4$. In such a case, it is possible that inflation happens after the flavor symmetry breaking but before the $Z_3$ breaking and hence the QCD anomaly effect may be enough to solve the domain wall problem~\cite{Riva:2010jm}.
}


\subsection{$D_4$ model}

A model based on the $D_4$ group can naturally explain the maximal atmospheric neutrino mixing angle~\cite{Grimus:2003kq,Ishimori:2008gp,Adulpravitchai:2008yp}.
An example of charge assignments under $D_4$ and $Z_N'$ for quarks, leptons and various Higgs fields are listed in Table~\ref{table:D4}.
We basically follow those of Ref.~\cite{Adulpravitchai:2008yp} where $N=5$ is assumed.

\begin{table}
\begin{center}
\begin{tabular}{|c|cccc|cccc|c|cccc|} \hline
    ~        & $Q_i$ & $u^c,c^c,t^c$ & $d^c$ &$s^c,b^c$ & $L_e$ & $L_I=(L_\mu,L_\tau)$ & $e^c$ & $e_I^c=(\mu^c,\tau^c)$  \\ \hline
 $D_4$   & ${\bf 1_{++}}$ & ${\bf 1_{++}}$ & ${\bf 1_{-+}}$ & ${\bf 1_{++}}$ & ${\bf 1_{++}}$ & ${\bf 2}$ & ${\bf 1_{++}}$ & ${\bf 2}$  \\ 
 $Z_N'$  &$1$ & $\omega_N^{-3}$ & $\omega_N$ & $\omega_N^{-1}$ & $\omega_N$ & $\omega_N$ & $1$ & $1$  \\ \hline
\end{tabular}
\vskip 3mm
\begin{tabular}{|c|cc|ccccc|} \hline
    ~        & $H_u$ & $H_d$ & $\chi_e$ & $\phi_e$ & $\chi_\nu$ & $\phi_\nu$ & $\xi_I=(\xi_1,\xi_2)$ \\ \hline
 $D_4$   & ${\bf 1_{++}}$ & ${\bf 1_{++}}$ & ${\bf 1_{++}}$ & ${\bf 1_{-+}}$ & ${\bf 1_{++}}$ & ${\bf 1_{+-}}$ & ${\bf 2}$  \\ 
 $Z_N'$  &$\omega_N^3$ & $\omega_N$ & $\omega_N^{-2}$ & $\omega_N^{-2}$ & $\omega_N^{-8}$ & $\omega_N^{-8}$ & $\omega_N^{-8}$  \\ \hline
\end{tabular}
\caption{Charge assignments under $D_4$ and $Z_N'$ for quarks, leptons and various Higgs fields. Here $\omega_N = e^{2\pi i /N}$ is the element of $Z_N'$ rotation.}
\label{table:D4}
\end{center}
\end{table}

The Yukawa interactions responsible for lepton mass matrices, up to the dimension six operators, are given as
\begin{align}
	\mathcal L = &y_{e1} \frac{e^c L_e H_d \chi_e}{M} + y_{e2} \frac{(e^c_I L_I)_{++} H_d \chi_e}{M} + y_{e3} \frac{(e^c_I L_I)_{-+} H_d \phi_e}{M} \nonumber\\
	&+ y_{\nu1}\frac{L_eL_eH_uH_u \chi_\nu}{M^2} + y_{\nu2} \frac{(L_I \xi_I)_{++} L_e H_u H_u}{M^2} \nonumber\\ 
	&+ y_{\nu3}\frac{(L_IL_I)_{++}H_uH_u \chi_\nu}{M^2} + y_{\nu4}\frac{(L_IL_I)_{+-}H_uH_u \phi_\nu}{M^2} + {\rm h.c.},
\end{align}
where $(\cdots)_{\pm\pm}$ denotes ${\bf 1_{\pm\pm}}$ product of $D_4$ doublets.
After having VEVs of
\begin{align}
	\left<\xi_I\right> = v_\xi(1,1),~~~
	\left<\chi_e\right> = v_{\chi_e},~~~\left<\phi_e\right>=v_{\phi_e},~~~
	\left<\chi_\nu\right> = v_{\chi_\nu},~~~\left<\phi_\nu\right>=v_{\phi_\nu},  \label{D4_min}
\end{align}
we will obtain desired charged lepton masses, neutrino masses, and
mixings with maximal atmospheric mixing angle and vanishing
$\theta_{13}$.  Higher order corrections modify this prediction and
non-zero $\theta_{13}$ can be obtained~\cite{Adulpravitchai:2008yp}.
These corrections do not modify the following argument at all.  The
only important fact is that there are $8N$ ($N=$ odd) or $4N$ ($N=$
even) discrete degenerate vacua connected by $D_4$ and $Z_N'$
transformations in this model.\footnote{
	As for the $Z_N'$ charges of two Higgs doublets, only the sum of
	$H_u$ and $H_d$ charges is relevant because of the degree of
	freedom of the hypercharge U(1) rotation.
}

As for the quark sector, we can write the Lagrangian as follows:
\begin{align}
	\mathcal L = y^{Ij}_d d^c_I Q_j H_d + y^{ij}_u  u^c_i Q_j H_u 
	+ y_d^{1j}\frac{d^c Q_j H_d \phi_e}{M} + {\rm h.c.},
\end{align}
where $i,j=1,2,3$ and $I=2,3$. After having VEVs of $\phi_e$, it becomes
generic Yukawa matrices for up and down type quarks, so it can clearly
reproduce the observed quark masses and mixings by choosing parameters
appropriately.  Since ${\bf 1_{-+}}$ changes its sign under both the
$Z_4$ and $Z_2$ rotation, which are subgroups of the $D_4$, they are
explicitly broken by the QCD anomaly and similarly $Z_N'$ is also
broken.  The anomaly coefficient (\ref{anom_coef}) becomes
\begin{align}
	\mathcal A = k_1\frac{1}{2} + k_2\frac{1}{2} + k_3 \frac{1}{N},
\end{align}
where $k_1=0,\dots,3$, $k_2=0,1$, and $k_3=0,\dots,N-1$ correspond to
$Z_4 \subset D_4$, $Z_2\subset D_4$, and $Z_N'$ discrete rotations,
respectively. Unfortunately, there are non-trivial cases that make
$\mathcal A$ integer for any choice of $N$: $(k_1,k_2,k_3)=(2,0,0),
(1,1,0)$.  The former means that the $Z_2$ rotation generated by the
group element $A^2$ (see App.~\ref{app:D4}) remains unbroken and the
latter means that another $Z_2$ rotation $AB (= BA^3)$ remains
unbroken. The latter $Z_2$ may be made anomalous if there exists another
strong gauge interaction,\footnote
{
In this case, it is possible that $Z_4 \subset D_4$ is broken by the QCD
anomaly, while $Z_2 \subset D_4$ is broken by the hidden SU$(N)$
anomaly or vice versa.  Then there only remains an unbroken $Z_2$
symmetry generated by the group element $A^2$.
}
but the former $Z_2$ always remain unbroken
independently of the details of the model, since there is no
non-trivial one-dimensional representation that is rotated by the angle
$e^{\pm i\pi/2}$ under $A$.  It means that the QCD anomaly does not
completely break the flavor symmetry but there remains at least one
unbroken $Z_2$ symmetry.  Thus domain walls exist in this class of
models.

\section{Conclusions} \label{sec:conc}

We have considered the effect of the QCD anomaly on the structure of
discrete vacua in models with discrete flavor symmetry assuming that some
of the quarks are charged under the discrete group.  It is found that
the anomaly only partially removes the vacuum degeneracy and cannot
remove all the degeneracy as far as the discrete groups listed in
Ref.~\cite{Ishimori:2010au} are considered.  It means that the domain
wall problem is not solved solely by the QCD anomaly effect.

Several remarks are in order.  The above conclusion was derived under
the assumption that the flavon VEVs spontaneously break the discrete
symmetry $D$ completely. It might be possible that flavon VEVs leave
some part of the original discrete symmetry $D$ unbroken. Then, if the
QCD anomaly explicitly breaks a part of the symmetry that will be
spontaneously ``broken'', domain walls become unstable.  For example,
let us suppose that $D=A_4$ is spontaneously broken down to $Z_2\times
Z_2$ by the flavon VEVs.  The domain walls associated with the
spontaneous breaking of $Z_3$ can be unstable since $Z_3$ can be
anomalous due to fermions in ${\bf 1'}$ or ${\bf 1''}$ representation.
It is non-trivial whether such a partial breaking of some discrete
symmetry could lead to phenomenologically interesting predictions for
quark/lepton masses and mixings. Furthermore, even if such a model-building will be successful,
we may need a relatively low scale of the flavor symmetry breaking, in order for the domain walls to collapse
due to the QCD anomaly effect before the domain wall domination of the universe.
It may be worth pursuing this possibility further.

We also remark that the existence of degenerate vacua does not
necessarily cause the cosmological domain wall problem.  As noted in
Introduction, if both the inflation scale and the reheating temperature
are sufficiently low, the flavor symmetry may never be restored after
inflation and there do not appear domain walls in the observable universe.\footnote{
	Inflation models in which the flavon takes a role of inflaton have been proposed~\cite{Antusch:2008gw,Antusch:2013toa,Ema:2016ops}.
}
The realistic situation can be more involved. In most known models of
the discrete flavor symmetry, there are several flavons that make the
whole scalar potential complicated and the VEV or mass of each flavon
can take hierarchically different value.  It means that the flavor
symmetry breaking scale may not necessarily be parametrized by just one
scale, and it can happen that only some part of the flavor symmetry
breaking occurs after inflation, and so on~\cite{Riva:2010jm}.  It should also be noticed
that we may put explicit symmetry breaking terms by hand at the
classical level unless the discrete symmetry is a remnant of some gauge
symmetry. Although it modifies the original prediction for the flavor
structure, the correction might be small enough to be neglected
phenomenologically while it can serve as a bias to make domain walls
unstable.

\section*{Acknowledgments}

We would like to thank Y.~Shimizu for useful discussion.  This work
was supported by the Grant-in-Aid for Scientific Research C (No.\
18K03609 [KN]), and Innovative Areas (No.\ 26104009 [KN], No.\
15H05888 [KN], No.\ 17H06359 [KN]).  This work was also supported by
JSPS KAKENHI Grant (No.\ 17J00813 [SC]).

\appendix

\section{Notes on discrete groups} \label{app}

\subsection{$A_4$ group} \label{app:A4}

$A_4$ is isomorphic to $(Z_2\times Z_2)\rtimes Z_3$ and it has 12 group elements. There are four representations (and hence four conjugacy classes), three of which are one-dimensional ${\bf 1}, {\bf 1'}, {\bf 1''}$ and the other is three-dimensional ${\bf 3}$.
Here we summarize explicit matrix form of the three-dimensional representation of the $A_4$ group elements~\cite{Altarelli:2010gt}:
\begin{align}
	&1 = \begin{pmatrix}
		1 & 0 & 0 \\
		0 & 1 & 0 \\
		0 & 0 & 1
	\end{pmatrix},~~~
	T= \begin{pmatrix}
		1 & 0 & 0 \\
		0 & \omega & 0 \\
		0 & 0 & \omega^2
	\end{pmatrix},~~~
	T^2=\begin{pmatrix}
		1 & 0 & 0 \\
		0 & \omega^2 & 0 \\
		0 & 0 & \omega
	\end{pmatrix}, \\
	&S = \frac{1}{3}\begin{pmatrix}
		-1 & 2 & 2 \\
		2 & -1 & 2 \\
		2 & 2 & -1
	\end{pmatrix},~
	ST= \frac{1}{3}\begin{pmatrix}
		-1 & 2\omega & 2\omega^2 \\
		2 & -\omega & 2\omega^2 \\
		2 & 2\omega & -\omega^2
	\end{pmatrix},~
	ST^2=\frac{1}{3}\begin{pmatrix}
		-1 & 2\omega^2 & 2\omega \\
		2 & -\omega^2 & 2\omega \\
		2 & 2\omega^2 & -\omega
	\end{pmatrix},\\
	&TS = \frac{1}{3}\begin{pmatrix}
		-1 & 2 & 2 \\
		2\omega & -\omega & 2\omega \\
		2\omega^2 & 2\omega^2 & -\omega^2
	\end{pmatrix},~
	TST= \frac{1}{3}\begin{pmatrix}
		-1 & 2\omega & 2\omega^2 \\
		2\omega & -\omega^2 & 2 \\
		2\omega^2 & 2 & -\omega
	\end{pmatrix},~
	TST^2=\frac{1}{3}\begin{pmatrix}
		-1 & 2\omega^2 & 2\omega \\
		2\omega & -1 & 2\omega^2 \\
		2\omega^2 & 2\omega & -1
	\end{pmatrix},\\
	&T^2S = \frac{1}{3}\begin{pmatrix}
		-1 & 2 & 2 \\
		2\omega^2 & -\omega^2 & 2\omega^2 \\
		2\omega & 2\omega & -\omega
	\end{pmatrix},~
	T^2ST= \frac{1}{3}\begin{pmatrix}
		-1 & 2\omega & 2\omega^2 \\
		2\omega^2 & -1 & 2\omega \\
		2\omega & 2\omega^2 & -1
	\end{pmatrix},~
	T^2ST^2=\frac{1}{3}\begin{pmatrix}
		-1 & 2\omega^2 & 2\omega \\
		2\omega^2 & -\omega & 2 \\
		2\omega & 2 & -\omega^2
	\end{pmatrix},
\end{align}
where $\omega = e^{2\pi i /3}$. Note that $T^2ST^2 = STS$.
The product of two $A_4$ triplets is decomposed as ${\bf 3}\times {\bf 3} = {\bf 1}+ {\bf 1'}+ {\bf 1''}+ {\bf 3}_S+ {\bf 3}_A$.
Explicitly, for two triplets $a=(a_1,a_2,a_3)$ and $b=(b_1,b_2,b_3)$, we find
\begin{align}
	&{\bf 1} \sim a_1b_1 + a_2b_3 + a_3b_2 \equiv (ab),\\
	&{\bf 1}'\sim a_3b_3 + a_1b_2 + a_2b_1 \equiv (ab)',\\
	&{\bf 1}''\sim a_2b_2 + a_3b_1 + a_1b_3 \equiv (ab)'',\\
	&{\bf 3}_S \sim \begin{pmatrix}
		2a_1b_1-a_2b_3-a_3b_2 \\
		2a_3b_3-a_1b_2-a_2b_1 \\
		2a_2b_2-a_3b_1-a_1b_3
	\end{pmatrix},~~~
	{\bf 3}_A \sim \begin{pmatrix}
		a_2b_3-a_3b_2 \\
		a_1b_2-a_2b_1 \\
		a_3b_1-a_1b_3
	\end{pmatrix}.
\end{align}

\subsection{$D_4$ group} \label{app:D4}

$D_4$ is isomorphic to $Z_4\rtimes Z_2$ and it has 8 group elements. There are five representations (and hence five conjugacy classes), four of which are one-dimensional ${\bf 1}_{++},{\bf 1}_{--},{\bf 1}_{+-},{\bf 1}_{-+}$ and the other is two-dimensional ${\bf 2}$.
Two-dimensional representation of the $D_4$ group elements are given by
\begin{align}
	&1 = \begin{pmatrix}
		1 & 0 \\
		0 & 1 
	\end{pmatrix},~~
	A = \begin{pmatrix}
		i & 0 \\
		0 & -i 
	\end{pmatrix},~~
	A^2 = \begin{pmatrix}
		-1 & 0 \\
		0 & -1 
	\end{pmatrix},~~
	A^3 = \begin{pmatrix}
		-i & 0 \\
		0 & i 
	\end{pmatrix},\\
	&B = \begin{pmatrix}
		0 & 1 \\
		1 & 0 
	\end{pmatrix},~~
	BA = \begin{pmatrix}
		0 & -i \\
		i & 0 
	\end{pmatrix},~~
	BA^2 = \begin{pmatrix}
		0 & -1 \\
		-1 & 0 
	\end{pmatrix},~~
	BA^3 = \begin{pmatrix}
		0 & i \\
		-i & 0 
	\end{pmatrix}.
\end{align}
Note that $ABA = B$.
The product of two $D_4$ doublets is decomposed as ${\bf 2}\times {\bf 2} = {\bf 1_{++}}+ {\bf 1_{--}}+ {\bf 1_{+-}}+ {\bf 1_{-+}}$.
Explicitly, for two doublets $a=(a_1,a_2)$ and $b=(b_1,b_2)$, we find
\begin{align}
	(ab)_{++} = a_1 b_2 + a_2 b_1,~~(ab)_{--} = a_1 b_2 - a_2 b_1,~~
	(ab)_{+-} = a_1 b_1 + a_2 b_2,~~(ab)_{-+} = a_1 b_1 - a_2 b_2.
\end{align}



\end{document}